\documentclass[twocolumn,superscriptaddress,amsmath,amssymb]{revtex4}
\usepackage{graphicx}% Include figure files
\usepackage{dcolumn}% Align table columns on decimal point
\usepackage{bm}% bold math
\usepackage{color}

\usepackage[colorlinks,
           bookmarks=true,
           linkcolor=blue,
           urlcolor=blue,
            anchorcolor=black,
            citecolor=blue
            ]{hyperref}

\usepackage[all]{hypcap}

\begin{document}

\title{Probing QCD critical fluctuations from intermittency analysis in relativistic heavy-ion collisions}

\author{Jin Wu, Yufu Lin, Yuanfang Wu, and Zhiming Li}
\email{lizm@mail.ccnu.edu.cn}
\affiliation{Key Laboratory of Quark and Lepton Physics (MOE) and Institute of Particle Physics, \\
Central China Normal University, Wuhan 430079, China}

\begin{abstract}
It is shown that intermittency, a self-similar correlation with respect to the size of the phase space volume, is sensitive to critical density fluctuations of baryon numbers in a system belonging to the three-dimensional (3D) Ising universality class. The relation between intermittency index and relative baryon density fluctuation is obtained. We thus suggest that measuring the intermittency in relativistic heavy-ion collisions could be used as a good probe of density fluctuations associated with the QCD critical phenomena. From recent preliminary results on neutron density fluctuations in central Au + Au collisions at $\sqrt{s_{NN}}$ = 7.7, 11.5, 19.6, 27, 39, 62.4 and 200 GeV at RHIC/STAR, the collision energy dependence of intermittency index is extracted and shows a non-monotonic behavior with a peak at around 20 - 27 GeV, indicating that the strength of intermittency becomes the largest in this energy region. The transport UrQMD model  without implementing critical physics cannot describe the observed behavior.

\end{abstract}

\maketitle
%%%%%%%%%%%%%%%%%%%%%%%%%%%%%%%%%%%%%%%%%%%%%%%%%%%%%%%%%%%%%%%%%%%%%%%%%%%%%%%
\section{Introduction}
The exploration of the QCD phase diagram and phase boundary is one of the main goals in relativistic heavy-ion collisions~\cite{StephanovPD,adams2005experimental,conservecharge0,conservecharge1}. At vanishing baryon chemical potential $\mu_B = 0$, finite temperature Lattice QCD calculations predict that a crossover from hadronic phase to the Quark Gluon Plasma (QGP) phase will occur~\cite{Lattice, Crossover}. QCD based model calculations indicate that the transition could be a first-order at large $\mu_B$~\cite{firstorder}. The point where the first-order phase transition ends is the so-called critical endpoint (CEP)~\cite{CEP1, CEP2}. Attempts such as searching for non-monotonic behaviors in the fluctuations of conserved quantities are being made to locate the CEP both theoretically~\cite{searchCEP1,searchCEP2,searchCEP3,searchCEP4} and experimentally~\cite{net_proton2010,net_proton2014,net_charge2014,net-kaon,phenix,luo2015energy,xiaofenglongpaper}. However, there exist sign problems in the Lattice calculations and large statistical uncertainties in the experimental measurements. Therefore solid conclusions on the existence and the location of the CEP cannot be reached at this stage. Further investigations from experiments, such as the Beam Energy Scan (BES) II program at RHIC~\cite{besII,besII2}, and precise Lattice QCD calculations or theoretical modeling are required.

Recently, it is argued that in heavy-ion collisions the baryon density fluctuations may provide a unique signal to the phase transition in the QCD phase diagram~\cite{DFphaseT1,DFphaseT2,DFphaseT3}. Critical opalescence~\cite{opale1,opale2,opale3}, which arises from the large fluctuations in the critical region in conventional QED matter, is a striking light scattering phenomenon happens in a continuous, or second-order, phase transition. In heavy-ion collisions, in analogy to the conventional critical opalescence, the created matter is expected to develop large baryon density fluctuations near the CEP due to a rapid increase correlation length in the critical region~\cite{StephanovPD,corrLen2}. There should exist strong fluctuations in the baryon density at kinetic freeze-out if such large density fluctuations can survive final-state interactions during the hadronic evolution of the system. The extracted neutron relative density fluctuation from light nuclei production in central Pb+Pb collisions at CERN SPS energies measured by NA49 Collaboration exhibits a peak structure around $\sqrt{s_{NN}}=8.8$ GeV, indicating that the CEP in the QCD phase diagram might have been reached or closely approached in this collision~\cite{DFphaseT2,DFphaseT3}.

In a grand canonical ensemble the correlation length diverges at the critical point. The system becomes scale invariant and fractality in multiparticle production processes~\cite{invariant1,invariant2,invariant3}. By using a critical equation of state for a 3D Ising system, which belongs to the same universality class of the QCD CEP~\cite{Ising1,Ising3,Ising4,Ising5}, it is found that the critical opalescence in the second-order phase transition shows as a strong intermittency behavior in QCD matter produced in high energy collisions~\cite{Antoniou2006PRL,Antoniou2010PRC,Antoniou2016PRC,Antoniou2018PRD}. Intermittency is a manifestation of the scale invariance and fractality of the physical process and the randomness of the underlying scaling law. It can be revealed in transverse momentum spectra as a pattern of power law behavior of scaled factorial moments~\cite{invariant1, Antoniou2006PRL}. The unique feature of this moment is that it can detect and characterize the non-statistical fluctuations in particle spectra, which are intimately connected with the dynamics of particle production. The power law fluctuations or intermittency behaviors have been recently observed in Si + Si collisions at 158A GeV/c from the NA49 experiment~\cite{NA49SFM} and in Ar + Sc collisions at 150A GeV/c from the NA61/SHINE experiment~\cite{NA61}.

Since baryon density fluctuations and intermittency properties are both sensitive to the QCD critical phenomena, an important question is that whether there exists a relationship between them. In this work, we plan to study this issue and try to get the relation between intermittency index and the relative baryon density fluctuation based on a critical Monte-Carlo model of the 3D Ising universality class. In experimental aspect, this could provide another effective method to estimate the baryon density fluctuation in relativistic heavy-ion collisions besides measuring the yield ratio of light nuclei productions~\cite{DFphaseT2,DFphaseT3}. From analyzing the preliminary experimental results on the neutron density fluctuations in central Au + Au collisions at STAR BES I energies~\cite{STARDF}, we obtain the energy dependence of the second-order intermittency index and compare it with the results from the 3D Ising model and the transport UrQMD model.

%%%%%%%%%%%%%%%%%%%%%%%%%%%%%%%%%%%%%%%%%%%%%%%%%%%%%%%%%%%%%%%%%%%%%%%%%%%%%%%
\section{A Critical Monte Carlo Model of the 3D Ising University Class}
For a 3D Ising universality system, the effective action can be written as~\cite{Antoniou2006PRL}:\\
\begin{equation}
	\Gamma_{c}[n_{B}]=T_{c}^{-5}g^{2}\int d^{3}\vec{x}[\frac{1}{2}|\nabla n_{B}|^{2}+Gg^{\delta-1}T_{c}^{8}|T_{c}^{-3}n_{B}|^{\delta+1}].
 \label{Eq:EA_Ising}
\end{equation}

\noindent Here, $n_{B}$ is the baryon-number density in the coordinate space, $\delta$ represents the isotherm critical exponent, $G$ means a coupling in the effective potential, and $g$ is a nonuniversal dimensionless constant.

Under circumstances that the contributions of $n_{B}$ to the partition function are boost invariant in the longitudinal direction, it has been proved~\cite{Antoniou2006PRL,Antoniou2016PRC} that the critical fluctuations and correlations developed in the coordinate space can be transferred to the momentum space for a small momentum transfer $\vec{k}$:\\
\begin{equation}
\lim\limits_{|\vec{k}|\rightarrow 0}\langle\rho_{\vec{k}}\rho_{\vec{k}}^{*}\rangle\sim |\vec{k}|^{-d_{F}},
 \label{Eq:CorrMom}
\end{equation}

\noindent where $\rho_{\vec{k}}$ is the Fourier transform of the baryon-number density from the coordinate space, and $\langle\rho_{\vec{k}}\rho_{\vec{k}}^{*}\rangle$ is the two-particle correlator in transverse momentum space. Eq.~\eqref{Eq:CorrMom} reveals a power law or fractal structure in momentum space with a fractal dimension $\tilde{d}_{F}=2-d_{F}$. This structure provides a tool for the detection of the critical point in heavy-ion experiments. If the QCD critical point belongs to the 3D Ising universality class~\cite{IsingCEP1,IsingCEP2}, the fractal dimension in transverse momentum space will be $\tilde{d}_{F}\simeq\frac{1}{3}$.

In high energy experiments, the power law or intermittency behaviors can be measured by calculations of scaled factorial moments (SFM) of final state particles. For this purpose, a region of chosen momentum space is partitioned into equal-size bins and the SFM is defined as:\\
\begin{equation}
F_{q}(M)=\frac{\langle\frac{1}{M^{D}}\sum_{i=1}^{M^{D}}n_{i}(n_{i}-1)\cdots(n_{i}-q+1)\rangle}{\langle\frac{1}{M^{D}}\sum_{i=1}^{M^{D}}n_{i}\rangle^{q}},
 \label{Eq:FM}
\end{equation}

\noindent with $M^{D}$ the number of cells or bins in which the D-dimensional momentum space is partitioned, $n_{i}$ the measured multiplicity in the $i$th bin, and $q$ the order of the moments.

A power law dependence of the SFM on the number of partitioned bins when $M$ is large enough indicates the presence of self-similar correlations in the system under investigation~\cite{SFM1,SFM2}:\\
\begin{equation}
F_{q}(M)\sim (M^{D})^{\phi_{q}}, M\rightarrow\infty.
 \label{Eq:PowerLaw}
\end{equation}

\noindent The index $\phi_{q}$, which characterizes the strength of the intermittency behavior, is called intermittency index. It has been shown to be related to the anomalous fractal dimension of the system~\cite{invariant3}. The study of multiplicity fluctuations in decreasing phase-space intervals using the method of SFM was first proposed several years ago~\cite{SFM1,SFM2}. Recent studies show that one can estimate the possible critical region of the QCD CEP by using the intermittency measurement together with the estimated freeze-out parameters~\cite{Antoniou2018PRD,CriticalReg}.

In order to study the intermittency behavior in detail by using the SFM method, we generate simulation events by implementing a critical Monte-Carlo (CMC) model~\cite{Antoniou2006PRL}. The simulation of CMC sample involving critical fluctuations in the baryon density requires the generation of baryon momenta correlated according to the power law of Eq.~\eqref{Eq:CorrMom}. A Levy random walk method~\cite{levy} is proposed to produce the momentum profile of the final state particles, with the probability density between two adjacent walks:\\
\begin{equation}
\rho(p)=\frac{\nu p_{\rm min}^{\nu}}{1-(p_{\rm min}/p_{\rm max})^{\nu}}p^{-1-\nu}.
 \label{Eq:probLevy}
\end{equation}

\noindent Here $p$ is the momentum distance of two particles which satisfying $p\in [p_{\rm min}, p_{\rm max}]$. The model parameters can be set to $\nu = 1/6$, $p_{\rm min} = 2\times 10^{-7}$ and $p_{\rm max} = 2$ for the 3D Ising universality class with the fractal dimension $\tilde{d}_{F}\simeq\frac{1}{3}$. The detailed description of the algorithm and implementation of the CMC model can be found in~\cite{Antoniou2006PRL,Antoniou2001NPA}.

%%%%%%%%%%%%%%%%%%%%%%%%%%%%%%%%%%%%%%%%%%%%%%%%%%%%%%%%%%%%%%%%%%%%%%%%%%%%%%%
\begin{figure}[!htb]
\hspace{-0.8cm}
\includegraphics[scale=0.36]{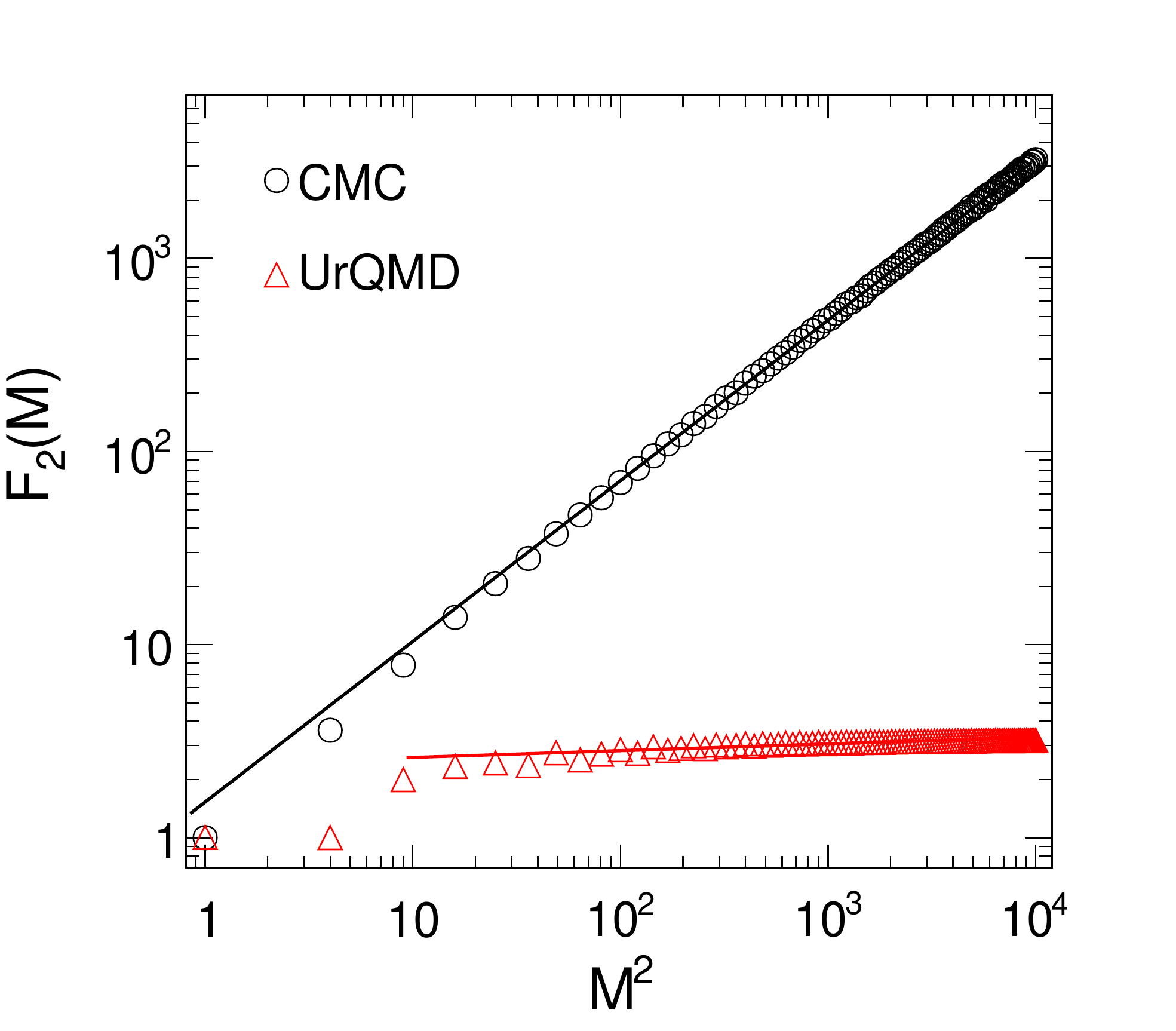}
\caption{The second-order SFMs as a function of number of partitioned bins in a double-logarithmic scale.}
\label{Fig:F2Scaling}
\end{figure}

For a system freeze out near the critical point, the correlation between particles distributed in a transverse momentum window at midrapidity region is supposed to be a self-similar structure~\cite{Antoniou2006PRL}. And the power-law behavior of Eq.~\eqref{Eq:PowerLaw} is expected to be held, with intermittency index $\phi_2=5/6$ determined by universality class arguments associated with the critical properties of a 3D Ising system~\cite{Antoniou2006PRL,NA49SFM}. In Fig.~\ref{Fig:F2Scaling}, the open black circles show the second-order SFMs as a function of the number of partitioned bins in a two-dimensional momentum space. The results are obtained for an ensemble of 600 critical events. In each event, the multiplicity distribution obeys a Poisson with the mean value $\langle n_B\rangle = 20$. The solid black line is a fitting according to Eq.~\eqref{Eq:PowerLaw}. It is clearly seen that the SFMs follow a good power law behavior with the increasing number of bins. It confirms that the CMC model can well reproduce the self-similar correlations as shown in Eq.~\eqref{Eq:CorrMom}. The fitting slope, i.e. the second-order intermittency index $\phi_2$, is found to be $0.834\pm0.001$, which is consistent with the theoretic expectation $\phi_2=5/6$ for a critical system with the fractal dimension $\tilde{d}_{F}\simeq\frac{1}{3}$~\cite{Antoniou2006PRL}. We have checked the systematic uncertainty and found it is less than 1\% by varying the model parameters $p_{\rm min}$ or $p_{\rm max}$.  The open red triangles in the same figure are results from the UrQMD model~\cite{UrQMD} with the same mean multiplicity around 20. The SFMs of the UrQMD model are found to be nearly flat with increasing number of cells, with the intermittency index is around 0. This is due to no critical related self-similar fluctuations implemented in the transport model.

%%%%%%%%%%%%%%%%%%%%%%%%%%%%%%%%%%%%%%%%%%%%%%%%%%%%%%%%%%%%%%%%%%%%%%%%%%%%%%%
\section{Baryon Density fluctuations and Self-similar Correlations}
%%%%%%%%%%%%%%%%%%%%%%%%%%%%%%%%%%%%%%%%%%%%%%%%%%%%%%%%%%%%%%%%%%%%%%%%%%%%%%%
\begin{figure}[!htb]
\hspace{-0.8cm}
\includegraphics[scale=0.36]{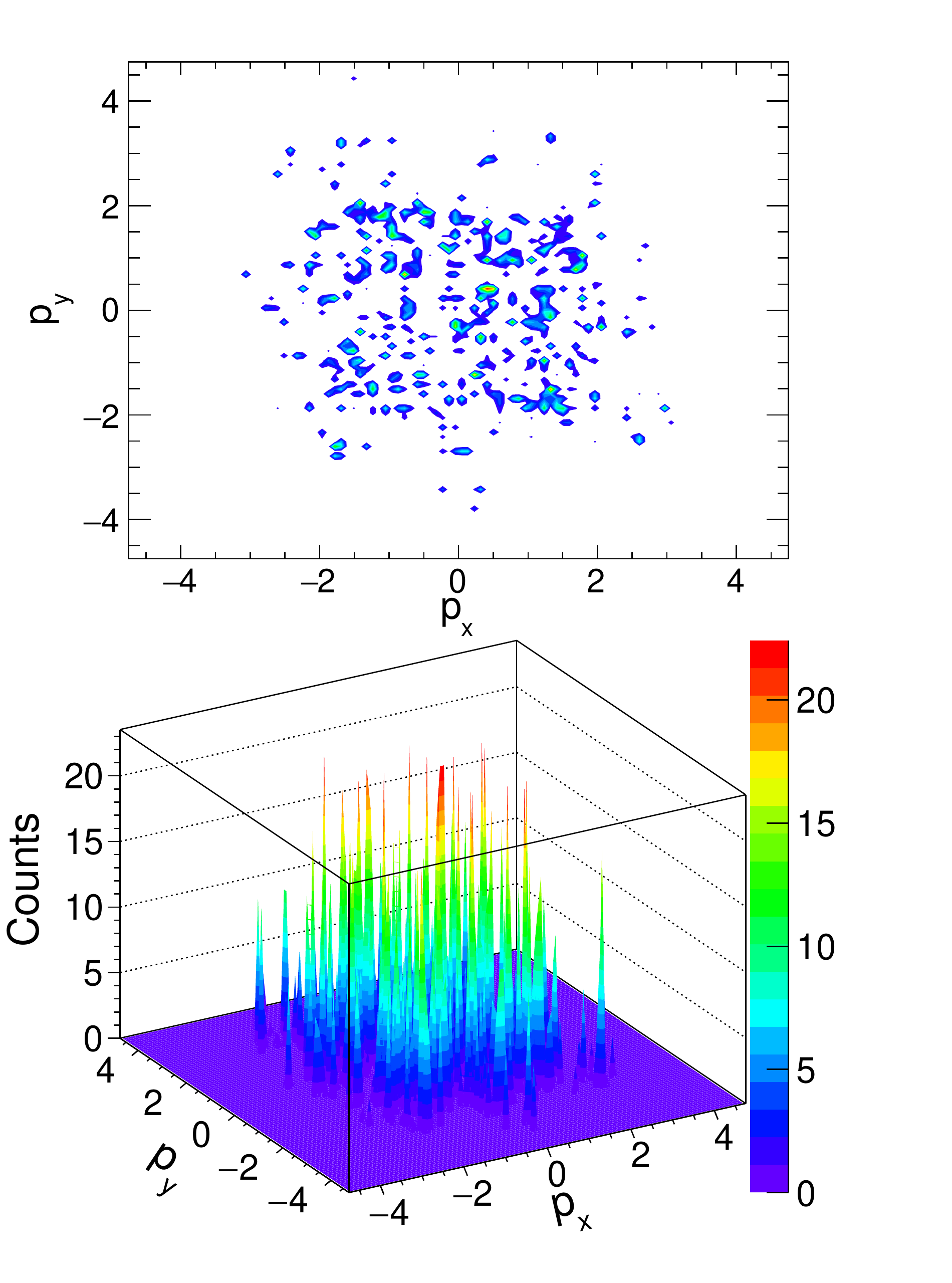}
\caption{Baryon density fluctuations in a 2D momentum space.}
\label{Fig:DensityFluc}
\end{figure}

In order to explore the baryon density fluctuations in the CMC model, we illustrate the density distribution in a 2D momentum space in the upper pad of Fig.~\ref{Fig:DensityFluc}. The lower pad shows the same plot with a contour view. From the figure, strong clustering effects in momentum space, which indicating giant phase-space density fluctuations are found. The observed large density fluctuations are probes of critical singularity of the system belong to the Ising universality class. This large local density fluctuation is suggested to be a manifestation of intermittency~\cite{Antoniou2006PRL}.

%%%%%%%%%%%%%%%%%%%%%%%%%%%%%%%%%%%%%%%%%%%%%%%%%%%%%%%%%%%%%%%%%%%%%%%%%%%%%%%
\begin{figure}[!htb]
\hspace{-0.8cm}
\includegraphics[scale=0.46]{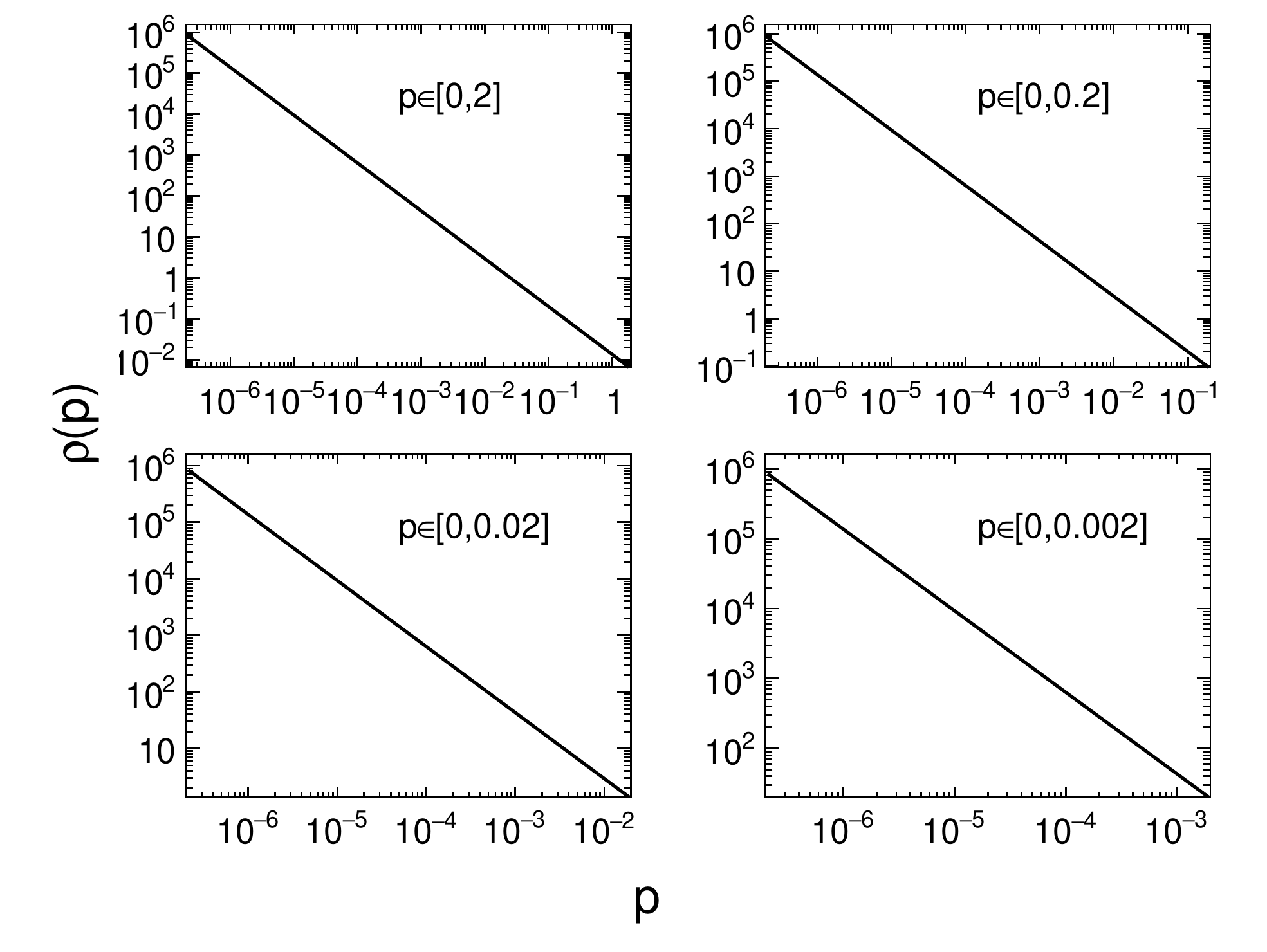}
\caption{The baryon density probability distribution in four different magnification scales.}
\label{Fig:SelfSimilar}
\end{figure}

It is argued that if intermittency occurs in particle production, large density fluctuations are not only expected, but should also exhibit a self-similarity behavior~\cite{invariant3}. In the current CMC model, the probability density distribution of two particles with distance $p$ in momentum space is given by Eq.~\eqref{Eq:probLevy}. It implies that two particle correlations are determined by the Levy distribution. The exponent of the Levy distribution is supposed to be related to the critical exponent of a system at a second-order phase transition~\cite{LevyExponent}. And it characterizes the power law structure of the particle correlation at the critical point~\cite{LevyPhenix}.

Figure \ref{Fig:SelfSimilar} presents the distributions of baryon density in four different magnification scales. We observe that the curves look the same at every level of magnification. They follow the same distribution in various momentum scales, i.e. scale invariant. Scale invariance is an exact form of self-similarity where at any magnification there is a smaller piece of the object that is similar to the whole. It is a typical character of a self-similar fractal system. The self-similar or intermittency nature of particle correlations in the CMC model is closely related to the large baryon density fluctuations which have been observed in Fig.~\ref{Fig:DensityFluc}, for a 3D Ising universality class system.

%%%%%%%%%%%%%%%%%%%%%%%%%%%%%%%%%%%%%%%%%%%%%%%%%%%%%%%%%%%%%%%%%%%%%%%%%%%%%%%
\section{Relation between Relative Baryon Density Fluctuation and Intermittency}
%%%%%%%%%%%%%%%%%%%%%%%%%%%%%%%%%%%%%%%%%%%%%%%%%%%%%%%%%%%%%%%%%%%%%%%%%%%%%%%
\begin{figure}[!htb]
\hspace{-0.8cm}
\includegraphics[scale=0.36]{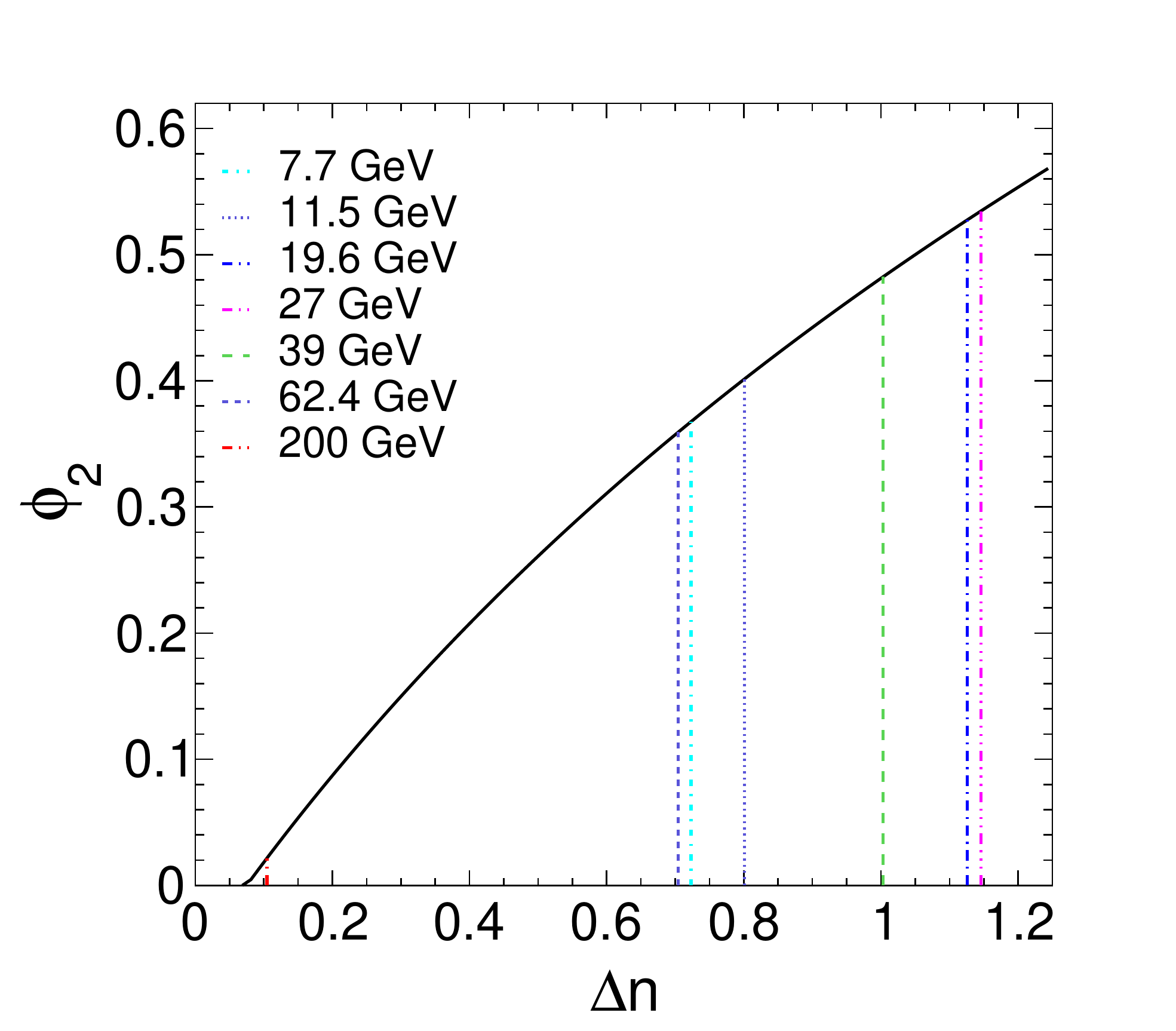}
\caption{The second-order intermittency index as a function of relative density fluctuation. The dash lines show the experimental measured relative neutron density fluctuations at RHIC/STAR~\cite{STARDF}.}
\label{Fig:RelationPhi2}
\end{figure}

%%%%%%%%%%%%%%%%%%%%%%%%%%%%%%%%%%%%%%%%%%%%%%%%%%%%%%%%%%%%%%%%%%%%%%%%%%%%%%%
In order to quantitatively describe density fluctuation, the relative density fluctuation of baryons $\Delta n$ is defined as ~\cite{DFphaseT2}:\\
\begin{equation}
\Delta n=\frac{\langle(\delta n)^2\rangle}{\langle n\rangle^2}=\frac{\langle n^2\rangle-\langle n\rangle^2}{\langle n\rangle^2}
 \label{Eq:relativeDF}
\end{equation}

\noindent where the angle bracket means the average in the phase space of the whole produced event sample.

By using the above introduced CMC algorithm method, we perform an event-by-event analysis on the scaled factorial moments and fit the intermittency index according to Eq.~\eqref{Eq:PowerLaw}. In the mean time, the baryon relative density fluctuations are also calculated from the produced baryons in the model. In Fig.~\ref{Fig:RelationPhi2}, the solid black line shows the second-order intermittency index $\phi_2$ as a function of $\Delta n$. The momentum range chosen in this analysis is $[-4.7, 4.7]$ for both $p_x$ and $p_y$. The $\phi_2$ is found to be monotonically increased with increasing relative density fluctuation $\Delta n$. Therefore, large intermittency is expected if giant baryon density fluctuations are developed near the QCD critical region. Furthermore, once the relation between $\Delta n$ and $\phi_2$ is obtained one can get the density fluctuations by measuring intermittency index from the same event sample, or vice versa. Thus, it provides an experimentally measurable quantity to estimate the density fluctuations in addition to measuring the light nuclei productions based on a coalescence model calculation~\cite{DFphaseT2,DFphaseT3}. Extracting the baryon density fluctuations in heavy-ion collisions from experimental observables is a challenging task as only the particle momentum distributions are generally measured. Alternatively, the relative neutron density fluctuations have been calculated from the measurement of the production yields of proton, deuteron, and triton in the central Au + Au collisions at RHIC/STAR~\cite{STARDF}. The dash lines in Fig.~\ref{Fig:RelationPhi2} display the values of measured neutron density fluctuations at $\sqrt{s_{NN}}$ = 7.7, 11.5, 19.6, 27, 39, 62.4 and 200 GeV, respectively.

%%%%%%%%%%%%%%%%%%%%%%%%%%%%%%%%%%%%%%%%%%%%%%%%%%%%%%%%%%%%%%%%%%%%%%%%%%%%%%%
\begin{figure}[!htb]
\hspace{-0.8cm}
\includegraphics[scale=0.36]{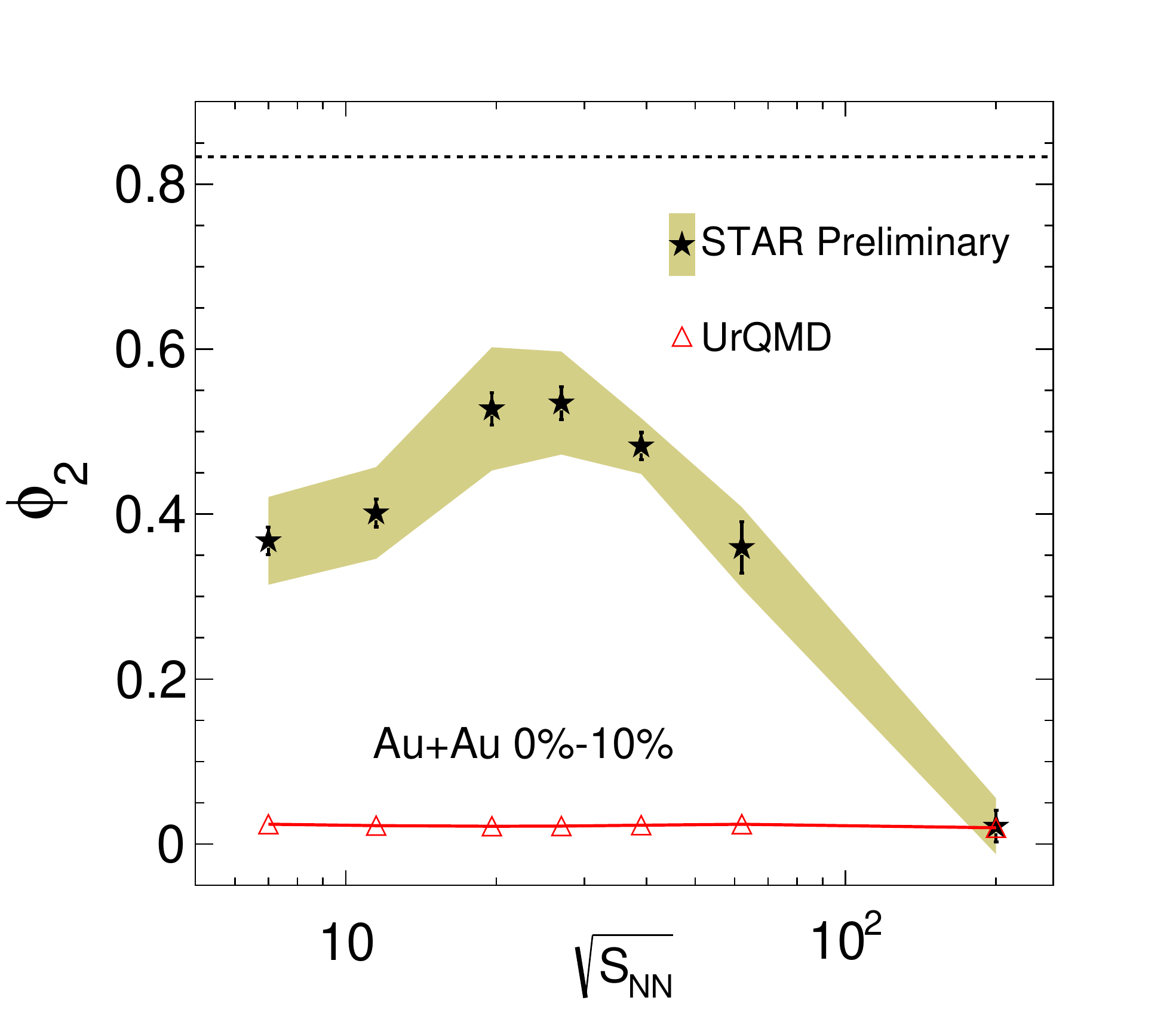}
\caption{Energy dependence of the second-order intermittency index in central Au + Au collisions at RHIC BES I energies.}
\label{Fig:STARPhi2}
\end{figure}

The black stars in Fig.~\ref{Fig:STARPhi2} show the energy dependence of the second-order intermittency index gained indirectly by mapping the STAR measured relative neutron density fluctuations in the most central Au + Au collisions (0\%-10\%) into the obtained relation between $\Delta n$ and $\phi_2$. The shadowed band illustrates the systematic errors from the experimental data. We observe that the strength of the intermittency first increases with energy, and then decreases rapidly in high energies with the value of $\phi_2$ approaching to zero at 200 GeV. The energy dependence of $\phi_2$ displays a non-monotonic behavior with a peak at around 20-27 GeV. In order to make a comparison, the black dash line presents the theoretic expectation of $\phi_2$ in the second-order phase transition from the 3D Ising model~\cite{Antoniou2006PRL}. One may notice that the values of the extracted $\phi_2$ from experimental measured density fluctuations are smaller than that from the Ising model. We should note that there are only seven measured data points available in experiment now. We need more data at RHIC BES II program to see whether it might touch the 3D Ising expectation at some energies between 20-27 GeV. The red symbols in the same figure show the UrQMD model calculations with the same kinematic cuts as the data. The UrQMD model is a transport model which does not include any CEP mechanisms~\cite{UrQMD}. Here we find that it shows a flat trend with the value around 0 at all energies.

We would like to point out that the results shown in this paper are based on an effective action belonging to the 3D Ising universality class. The boost invariant property of the baryon-number densities in the longitudinal momentum space is assumed. Although this assumption is valid at most of the RHIC energies in central rapidity regions, it still needs to be carefully checked at lower beam energies. Moreover, we use the neutron density fluctuation calculated from the STAR data to get Fig.~\ref{Fig:STARPhi2} since there is no experimental measured baryon density fluctuation in the market. Future experimental and theoretical investigations are need to verify the present results. In addition, we would note that  some effects including the global conservation of energy and momentum, late stage rescattering, detection efficiency and so on need to be taken into account in experimental measurement of these quantities to get a clean signature of the QCD critical point.

%%%%%%%%%%%%%%%%%%%%%%%%%%%%%%%%%%%%%%%%%%%%%%%%%%%%%%%%%%%%%%%%%%%%%%%%%%%%%%%
\section{Conclusions and Outlook}
%%%%%%%%%%%%%%%%%%%%%%%%%%%%%%%%%%%%%%%%%%%%%%%%%%%%%%%%%%%%%%%%%%%%%%%%%%%%%%%
In summary, with a 3D Ising universality class CMC model, we have demonstrated that large baryon density fluctuations are observed in a 2D momentum space. The self-similar or intermittency nature of particle correlations in the CMC model is closely related to the observed large baryon density fluctuations associated with the critical point. The relation between intermittency index and relative density fluctuation has been obtained, suggesting that the information on the relative density fluctuation of baryons can be determined from the measurement of intermittency of the system, or vice versa. The extracted intermittency index from the neutron density fluctuations measured by STAR in the central Au + Au collisions at RHIC BES I energies exhibits a non-monotonic energy dependence with a peak at around 20-27 GeV. The comparison of results between the extracted experimental data and the UrQMD calculations shows that the non-monotonic behavior cannot be described by the UrQMD model without implementing critical physics.

In the experimental exploration of the QCD phase diagram, corresponding observables to probe the signature of the QCD phase transition and CEP in heavy-ion collisions have been measured at RHIC BES I energies during the past years. In particularly, some intriguing non-monotonic behaviors in the Au + Au collisions have been found. In the dependence of collision energies, the higher cumulant ratios and the directed flow slope of protons have a dip shape~\cite{luo2015energy,v1STAR,v1STAR2} as well as the neutron density fluctuation and the HBT radii~\cite{STARDF,HBTSTAR} show a peak structure. All of these non-monotonic behaviors are found at energies around $20<\sqrt{s_{NN}}<30$ GeV. Here, we have shown that the intermittency index should also present a non-monotonic energy dependence given the phase transition of the QCD CEP belongs to the same universality class with the 3D Ising model. The energy dependence of intermittency index could be used as a good probe of large density fluctuations associated with the QCD critical phenomena.

The RHIC experiment has planned a second energy scan of BES II program to run in 2019 - 2020. With significant improved statistics and particle identification in BES II, it would be of great interests if STAR could directly measure the intermittency together with other sensitive observables to precisely determine the location of the CEP in the QCD phase diagram.
%%%%%%%%%%%%%%%%%%%%%%%%%%%%%%%%%%%%%%%%%%%%%%%%%%%%%%%%%%%%%%%%%%%%%%%%%%%%%%%
\section*{Acknowledgments}
%%%%%%%%%%%%%%%%%%%%%%%%%%%%%%%%%%%%%%%%%%%%%%%%%%%%%%%%%%%%%%%%%%%%%%%%%%%%%%%
We thank Dr. Nu Xu and Xiaofeng Luo for the fruitful discussions and comments. We further thank the STAR Collaboration for providing us with their preliminary data. This work is supported by the Ministry of Science and Technology (MOST) under Grant No. 2016YFE0104800 and the Fundamental Research Funds for the Central Universities under Grant No. CCNU19ZN019.  

%\bibliography{ref}% BiTex form
%\include{ref}
%%%%%%%%%%%%%%%%%% Citation %%%%%%%%%%%%%%%%%%%%%%%%%

\end{document}